\documentclass[12pt,epsf]{article}
\usepackage{graphicx, amsmath}

\begin{document}

\sloppy
\begin{flushright}{SIT-HEP/TM-44}
\end{flushright}
\vskip 1.5 truecm
\centerline{\large{\bf Modulated inflation from kinetic term}}
\vskip .75 truecm
\centerline{\bf Tomohiro Matsuda\footnote{matsuda@sit.ac.jp}}
\vskip .4 truecm
\centerline {\it Laboratory of Physics, Saitama Institute of Technology,}
\centerline {\it Fusaiji, Okabe-machi, Saitama 369-0293, 
Japan}
\vskip 1. truecm
\makeatletter
\@addtoreset{equation}{section}
\def\theequation{\thesection.\arabic{equation}}
\makeatother
\vskip 1. truecm

\begin{abstract}
\hspace*{\parindent}
We study modulated inflation from kinetic term. 
Using the Mukhanov-Sasaki variable, it is possible to determine
 how mixing induced by the kinetic term feeds the curvature
 perturbation with the isocurvature perturbation. 
We show explicitly that the analytic result obtained from the evolution
 of the Mukhanov-Sasaki variable is consistent with the 
$\delta N$-formula.  
From our results, we find analytic conditions for the modulated
 fluctuation and the non-Gaussianity parameter. 
\end{abstract}

\newpage
\section{Introduction}
Since the mass of the Higgs field in the Standard Model (SM) is much
smaller than the Planck scale, it is natural to expect that there is a
mechanism (solution to the hierarchy problem) that causes the mass of
the scalar fields to be much lighter than the Planck scale. 
In fact, string theory and supersymmetric models predict many light
scalar fields whose expectation values determine the parameters of
low-energy effective action. 
During inflation, light fields (${\cal M}_i$) may lead to vacuum
fluctuations that appear as classical random Gaussian inhomogeneities
with an almost scale-free spectrum of amplitude $\delta {\cal M}_i$.
Since the wavelength of the fluctuations is stretched during inflation
over the Hubble horizon after inflation, the vacuum fluctuations of
the light fields can be related in various ways to the cosmological
curvature perturbation in the present Universe. In this paper, we
consider modulated inflation as a mechanism relating the isocurvature
perturbation to the curvature perturbation of the Universe. 
In this paper, we consider modulated inflation as one of the
mechanisms that relate the isocurvature perturbation 
to the curvature perturbation of the Universe.
The basic idea of modulated inflation is very simple. 
We first introduce modulated inflation 
mentioning the difference between modulated perturbation scenarios and
multi-field (double) inflation.  

We start with the conventional equation for the number of e-foldings
elapsed during inflation:
\begin{equation}
\label{original_1}
N =\frac{1}{M_p^2}\int^{\phi_N}_{\phi_e} \frac{V}{V_\phi}d\phi,
\end{equation}
where $\phi_N$ is the value of inflaton field $\phi$ corresponding to
$N$ e-foldings, and $\phi_e$ denotes the end-point of inflation.
Using the $\delta N$-formula, we find that the fluctuation of the
spectrum $\delta \phi_N = H_I/2\pi$ leads to the spectrum of the density
perturbation
\begin{equation}
 \delta_H^2 = \frac{4}{25}(\delta N)^2 
=\frac{4}{25}\left(\frac{V}{M_p^2V_\phi}\frac{H_I}{2\pi}\right)^2,
\end{equation}
where $M_p$ is the reduced Planck mass and $H_I$ is the Hubble parameter
during inflation.
In addition to the standard inflation scenario in which $\delta \phi_N$
leads to the density perturbation, one may expect more generically 
other scalar fields may play a similar role. 
The first specific example of this has been given by
Bernardeau et al.\cite{modulated-inf1} for modulated couplings in
hybrid-type inflation, in which $\phi_e$ depends on light fields
through moduli-dependent couplings.
Lyth\cite{modulated-inf1} considered a multi-inflaton model of
hybrid inflation and found another realization of 
$\delta \phi_e$-induced curvature perturbation:
``generating the curvature perturbation at the end of inflation''.
More recently, we considered trapping inflation combined with
inhomogeneous preheating and found a different
mechanism for generating the curvature perturbation at the end of
weak inflation\cite{alternative-PR} caused by the
fluctuation of the number density ($\delta n$) of the preheating field. 

In addition to modulated scenarios related to the perturbation
$\delta \phi_e$, we have recently proposed a new modulated scenario,
modulated inflation\cite{modulated-inflation}.  
In Eq.(\ref{original_1}), fluctuations induced by the other components
$V_\phi$ and $M_p$ may  
generate curvature perturbation if the
components are modulated during inflation.
Based on this simple idea, we considered a new
mechanism for generating the curvature
perturbation\cite{modulated-inflation} that relies 
on neither $\delta \phi_N$ nor $\delta \phi_e$.
The source of the curvature perturbation in this scenario is 
the explicit ${\cal M}$-dependence of the inflaton velocity  
$\dot{\phi}\simeq V_\phi/3H_I$.
The equation that relates the $\delta \dot{\phi}$-perturbation 
to the curvature perturbation is $N_\phi =H_I/\dot{\phi}$.
The integration $N=\int^{\phi_N}_{\phi_e}N_\phi d\phi$ allows
$\delta N_\phi$ to feed the curvature perturbation with
the modulated perturbation continuously.\footnote{See also Fig.1.}
Another way to see the source in the new scenario is to consider the
conventional evolution of the curvature perturbation   
\begin{equation}
\dot{\cal R}=-H\frac{\delta P}{\rho+P},
\end{equation}
where the pressure perturbation $\delta P\simeq \dot{\phi}\delta
\dot{\phi}$ is generated by the modulated perturbation.

\begin{figure}[h]
 \begin{center}
\begin{picture}(400,250)(0,0)
\resizebox{15cm}{!}{\includegraphics{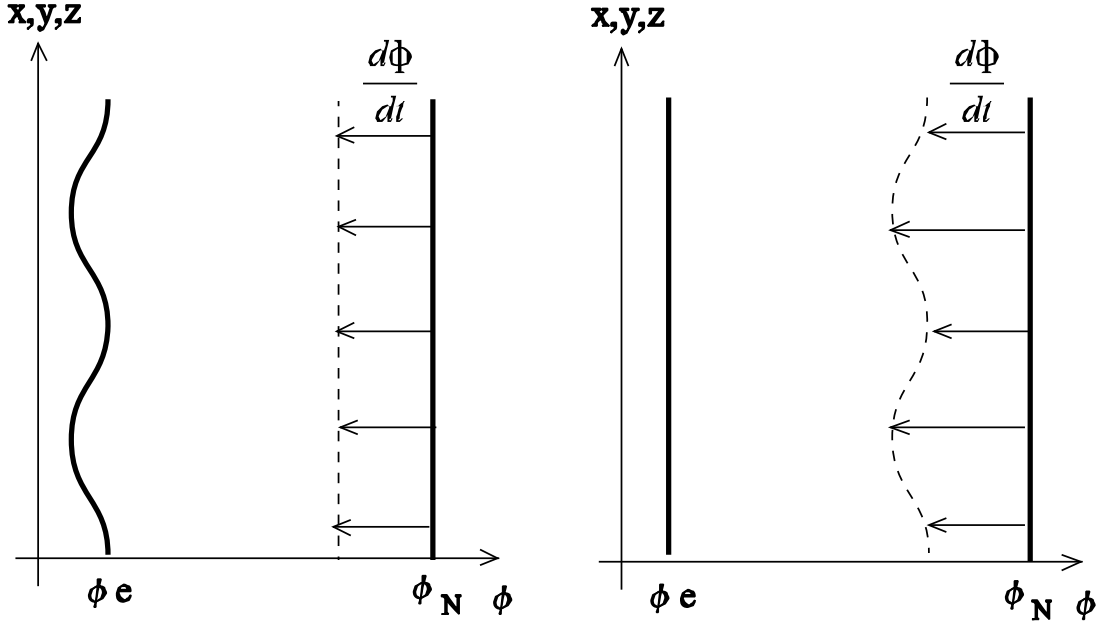}} 
\end{picture}
\caption{We consider the equation 
$\delta N =H \delta t_N$, where $t_N$ denotes the time elapsed 
during inflation (from $\phi=\phi_N$ till the end $\phi=\phi_e$). 
In the left, the curvature perturbation is generated due to the
modulated fluctuation $\delta \phi_e$ that leads to $\delta t_N$.
In the right, the fluctuation is related to the modulated fluctuation
$\delta \dot{\phi}$. } 
\label{fig1}
 \end{center}
\end{figure}

On the other hand, in multi-field (double) inflation
scenario\cite{two-field},   the
change in the co-moving curvature perturbation caused by the isocurvature
perturbation $\delta s$ is given by
\begin{equation}
\dot{\cal R}|_s\simeq\frac{2H_I}{\dot{\sigma}}\dot{\theta}\delta s,
\end{equation}
where $\sigma$ denotes the adiabatic component
 and $\dot{\theta}\ne 0$ appears at the ``bend'' in the inflaton
 trajectory.
This discriminates the source terms in modulated perturbation scenarios
 from multi-field (double) inflation.
The term ``modulated perturbation''\cite{modulated-inf1,
modulated-inflation} has been used to distinguish different origins of
 the cosmological perturbation. 
It might be confusing, but in a modulated perturbation scenario the
 light field (${\cal M}$) can be identified with an additional
 inflaton field.\footnote{For
 specific example, see Ref.\cite{modulated-inf1, alternative-PR,
alt-string}.}
In this case, there can be two different sources;
 ``bend'' and the modulated perturbations.
This approach is useful for brane inflation,
 since there can be several directions for the moving
brane, as well as for the target brane.

As we mentioned above, there is a crucial difference between
multi-field (double) inflation\cite{two-field} and modulated
perturbation scenarios\cite{modulated-inf1, modulated-inflation} 
in the mechanism that converts the isocurvature perturbation into
curvature perturbation. 
However, the difference is not obvious when a lght field appears in the
inflaton kinetic term.
In section 2, we examine the differences and the similarities
between the two scenarios when a lght field appears in the
inflaton kinetic term.

\section{Modulated inflation from kinetic term}
\hspace*{\parindent}
We introduce a light field ${\cal M}$ and consider the kinetic
term for the inflaton $\phi$;
\begin{equation}
S=\int d^4x\sqrt{-g}\left[\frac{1}{2\kappa^2}R
-\frac{1}{2}\left(\nabla{\cal M}\right)^2
-\frac{1}{2}\omega({\cal M})
\left(\nabla\phi\right)^2
-W(\phi,{\cal M})\right],
\end{equation}
where $\frac{\kappa^2}{8\pi}=G$ is the Newton's gravitational constant
and $\omega({\cal M})$ is a function of the moduli.
The kinetic term may be given by a
more general function, since any term that is
not forbidden by symmetry may appear in the effective action.
However, for the effective action during
inflation and the kinetic term that can be approximated
by a series expansion, we may disregard terms proportional to higher
$(\nabla \phi)^n$.
Of course, the above approximation is not always true, but we consider
such action so that we can follow Ref.\cite{Lalak-kinmix} in the
following.\footnote{See Ref.\cite{reviewer1} for more specific examples. 
In Ref.\cite{reviewer1}, Ringeval et al. considered multi-field
inflation in a brane model in which the specific form of the
kinetic term is obtained from the string theory.} 

We first consider a separable potential
\begin{equation}
W= V(\phi)+X({\cal M}),
\end{equation}
where $V(\phi)$ is the conventional inflaton potential.
We consider flat potential for the light field $\cal M$.
The coefficient of the kinetic term may be written as
$\omega({\cal M})=1+\beta \frac{{\cal M}^2}{M_p^2}$ or
$\omega({\cal M})= e^{b({\cal M})}$.\footnote{Note that the former is
identical to the latter when $b({\cal M})$ is given by a logarithmic
function. }
These terms may appear in low-energy effective action.
The definitions of the slow-roll parameters are
\begin{eqnarray}
\epsilon_{\cal M}&\equiv& \frac{M_p^2}{2}\left(\frac{X'}{W}\right)^2
\nonumber\\
\epsilon_{\phi}&\equiv& \frac{M_p^2}{2\omega}\left(\frac{V'}{W}\right)^2,
\nonumber\\
\epsilon &\equiv& \frac{\omega\dot{\phi}^2}{2M_p^2 H_I^2}+\frac{\dot{\cal M}^2}{2M_p^2 H_I^2},
\end{eqnarray}
where the prime denotes the derivative of the potential with respect to
the corresponding field.

Note that we are not considering double inflation in which the
isocurvature perturbation feeds the curvature perturbation
 mainly when there is a sharp bend in the trajectory.
The bend occurs if the additional inflation
stage is induced by the secondary inflaton field.
Instead, we consider modulated inflation, in which
modulated perturbation of the inflaton velocity sources
the curvature perturbation.
We first consider the evolution of the curvature
perturbation paying attention to the source terms that can feed the
curvature perturbation in different ways.

\subsection{Evolution of the curvature perturbation}
Recently, cosmological perturbation in two-field inflation with a
light field appearing in the inflaton kinetic term has been
studied by Lalak et al.\cite{Lalak-kinmix}.
They calculated the spectra of curvature and isocurvature modes at the
Hubble crossing and computed numerically the evolution of the curvature
and isocurvature modes, showing how isocurvature perturbations 
significantly affect
 the curvature perturbation after Hubble crossing.
Our first task is to obtain the analytic form of the curvature
perturbation generated by the constant source term that can be related
to the modulated inflation.
We mainly follow Ref.\cite{Lalak-kinmix} and consider
 the Mukhanov-Sasaki variable. 
At this time, we do not assume coincidence of the calculation given in
Ref.\cite{Lalak-kinmix} and the $\delta
N$-formula, since the source term appeared in the $\delta N$-formula
has been disregarded in the similar calculation when the inflaton has 
the standard kinetic term\cite{two-field}. 
As will be shown later in the paper, the calculation is
very simple in the $\delta N$-formalism.
For convenience, we follow the notations in
Ref.\cite{Lalak-kinmix}, with a slight difference.
In Ref.\cite{Lalak-kinmix}, the light field ${\cal M}$ and the inflaton
$\phi$ are denoted by $\phi$ and $\chi$, respectively, and the
coefficient $\omega({\cal M})$ is given by an exponential $\omega
=e^{b(\phi)}$. 
Since we are considering modulated inflation, the adiabatic
component $\sigma$ is essentially the same as the inflaton $\phi$.
Following the definitions in Ref.\cite{Lalak-kinmix}, we thus find for
modulated inflation that the trajectory is given by
\begin{eqnarray}
\cos \theta &\equiv& \frac{\dot{\cal M}}{\dot{\sigma}}\simeq 0,\nonumber\\
\sin \theta &\equiv& 
\frac{\dot{\phi}\omega^{1/2}}{\dot{\sigma}}\simeq 1, 
\end{eqnarray}
where $\theta$ is a constant in modulated inflation, but changes abruptly
when there is a sharp bend in double inflation.
We do not consider the case when there is a sharp bend.
The Mukhanov-Sasaki variable is defined by
\begin{equation}
Q_\sigma\equiv \delta \sigma + \frac{\dot{\sigma}}{H}\Phi,
\end{equation}
where $\Phi$ is the metric perturbation.
Using the slow-roll approximations, the equation of motion for the
perturbation gives\cite{Lalak-kinmix}
\begin{eqnarray}
\label{eqforQ}
\dot{Q_\sigma}&\simeq& AHQ_\sigma + BH \delta s\nonumber\\
\dot{\delta s} &\simeq& DH \delta s,
\end{eqnarray}
where $\delta s = \delta {\cal M}$ for modulated inflation.\footnote{Note
that the constant source term proportional to $H \delta s$  
has been disregarded in Ref.\cite{two-field}.
Therefore, the absence of this term discriminates
between modulated inflation\cite{modulated-inflation} and multi-field
inflation for standard kinetic terms, as discussed in the introduction.
On the other hand, the term proportional to $H \delta s$ appears in the
similar calculation when there is a light field appearing in the
inflaton kinetic term\cite{Lalak-kinmix}. 
In this paper we examine the validity of this approach and the meaning
of the term comparing with the $\delta N$-formalism.}
$A, B, D$ are given by the following formula:
\begin{eqnarray}
\label{ABD}
A &=&
-\eta_{\sigma\sigma}+2\epsilon - \xi \cos\theta \sin^2\theta
\simeq -\eta_{\sigma\sigma}+2\epsilon \nonumber\\
B &=& 
-2\eta_{\sigma s}+ 2\xi \sin^3\theta 
\simeq 2\frac{d\theta}{d N} -2\xi \sin\theta
\simeq -2\xi \nonumber\\
D &=& -\eta_{ss} + \xi \cos\theta(1+\sin^2\theta)
\simeq -\eta_{ss} \simeq 0,
\end{eqnarray}
where $\xi$ is defined by\footnote{For the definitions of the 
$\eta$-parameters, see Ref.\cite{Lalak-kinmix}. }
\begin{equation}
\xi \equiv \frac{1}{\sqrt{2}}\frac{\omega'}{\omega}M_p \sqrt{\epsilon}.
\end{equation}
The source of the isocurvature perturbation feeding appears
in Eq.(\ref{eqforQ}) in the term proportional to $B$.
In double inflation, there is a significant feeding with
isocurvature perturbation if there is a sharp bend in the
trajectory.
The sharp bend leads to a large $d\theta/d N$, as has been 
discussed for double inflation\cite{Bernardeau2002jy}.
In our scenario of modulated inflation, we do not
expect such a bend in the trajectory.
The main source in modulated inflation is the term
proportional to $\xi$, which is small compared with $d\theta/d N$ at
the bend, but gives a constant feeding and may become significant after 
integration.
Because of the integration, the correction will be proportional to the
number of e-foldings.
Disregarding the first term in Eq.(\ref{eqforQ}), we find 
\begin{equation}
\Delta Q\sim \int BH \delta s dt \sim B \delta s \int H dt\sim
 -2\xi N \delta s ,
\end{equation}
which gives a simple result for the co-moving curvature perturbation:
\begin{equation}
\Delta{\cal R}\equiv \frac{H}{\dot{\sigma}}\Delta Q_\sigma \sim 
-\frac{H}{\dot{\sigma}} 
\left[\sqrt{2}\frac{\omega'}{\omega}M_p \sqrt{\epsilon}\delta s N\right]
\simeq -sign(\dot{\sigma})\frac{\omega'}{\omega}N \delta s.
\end{equation}
These approximations are valid only if the time-dependence
of the slow-roll parameters is small between Hubble crossing and
the end of inflation.
In the next subsection, we examine the slow-roll conditions 
used in the calculation\cite{Lalak-kinmix}.

Note that an inflation model with vanishing $Q_\sigma$ at the horizon
crossing does not cause a serious problem in modulated inflation
scenario. 
Modulated inflation involves continuous feeding with
isocurvature perturbation, raising the curvature perturbation after
horizon crossing.
The feeding mechanism operates even if the
standard inflaton perturbation is very small at the horizon crossing.

\subsection{$\delta N$-formalism}
Our second task is to find analytic result for the curvature
 perturbation following the $\delta N$-formalism.
Variation of the action leads to the equations\cite{Mukhanov:1997fw}
\begin{eqnarray}
\ddot{\phi}+3H_I\dot{\phi}+ 
\frac{V'}{\omega}+\frac{\omega'}{\omega}\dot{\phi}\dot{\cal M}&=&0
\nonumber\\
\ddot{\cal M}+3H_I\dot{\cal M}+ X'
-\frac{1}{2}\omega'\dot{\phi}^2&=&0.
\end{eqnarray}
Following Ref.\cite{Lalak-kinmix}, we find the velocity for the
slow-roll inflaton field to be
\begin{equation}
\label{infla-appr}
\dot{\phi}\simeq -\frac{V'}{3H_I\omega},  
\end{equation}
where the approximation is valid if 
$|\omega' \dot{\cal M}/(3H_I \omega)| \ll 1$.
Because of the additional term proportional to $\dot{\phi}^2$, the
slow-roll condition for the light field ${\cal M}$ is 
\begin{equation}
\frac{M_p^2}{2}\left(\frac{X'-\omega'\dot{\phi}^2/2}{W}\right)^2 \ll 1.
\end{equation}
If the potential $X$ for the field ${\cal M}$ is very flat ($X'\simeq
0$), the above condition leads to
\begin{equation}
\frac{|\omega'|}{\omega}\ll \frac{3\sqrt{2}}{M_p \epsilon_\phi}.
\end{equation}
This is not the condition for the potential, but is needed to ensure
the slow motion of the field ${\cal M}$ during inflation.
Otherwise, we find the conventional slow-roll condition
$\epsilon_{\cal M}\ll 1$ when the term proportional to $\dot{\phi}$ is
negligible compared with $X'$.

In modulated inflation scenario, the analytic calculation 
 is very simple in the $\delta N$-formalism\cite{modulated-inflation}. 
We follow Ref.\cite{modulated-inflation} and start with the following
formula\cite{delta-N} 
\begin{equation}
N_{\phi}\dot{\phi}=-H_I,
\end{equation}
where the subscript denotes the derivative with respect to the
corresponding field.
Using Eq.(\ref{infla-appr}), we find 
\begin{equation}
N_\phi = -\frac{H_I}{\dot{\phi}}\simeq \frac{3H_I^2}{V'}\omega.
\end{equation}
The meaning of this equation is clear.
Considering $\phi(N)$ as the ``time'' during inflation,
$N_\phi$ gives the rate of change in the number of e-foldings.
If $N_\phi$ is perturbed by the modulated fluctuation 
($\delta \omega \ne0$), it leads to $\delta N$ after the
``time''-integration.  
If $\omega$ is a constant during inflation, we find
\begin{equation}
N\simeq \omega \int \frac{3H_I^2}{V'} d\phi.
\end{equation}
The fluctuation of the light field $\delta {\cal M}$ thus gives 
\begin{equation}
\delta N \simeq \frac{\omega'}\omega N \delta {\cal M}.
\end{equation}
The $\delta N$-formula is a very powerful tool in calculating
the curvature perturbation in this set-up, since we
can obtain analytic result without paying special attention to the
curvature-isocurvature mixing during inflation. 
The fact that the two different calculations lead to the identical
result is an important finding, since for standard kinetic terms there
was no such correpondence.

Finally, we examine the conditions related to the cosmic microwave
background (CMB) spectra. 
The condition for the modulated perturbation to dominate
the curvature perturbation is
\begin{equation}
\label{result1}
\left|\frac{N \omega'\omega^{-1}}{\sqrt{1/2\epsilon}M_p^{-1}} 
\frac{\delta {\cal M}}{\delta \sigma}\right|
\simeq  \left|N \frac{\omega'}{\omega} M_p\epsilon^{1/2} 
\frac{\delta {\cal M}}{\delta \sigma}\right| >1.
\end{equation}
If the modulated perturbation dominates curvature 
perturbation, the non-Gaussianity parameter can be
large.\footnote{Non-Gaussian metric fluctuations related to the
fluctuation $\delta \phi_e$ have been discussed by Bernardeau et al. in
Ref.\cite{Bernardeau:2002jf}, prior to the publication of 
Ref.\cite{modulated-inf1}. Note that in modulated inflation the source
of the curvature perturbation is not related to $\delta \phi_e$.}
Note that the origin of this parameter is different from the that
obtained in double inflation with a sharp bend\cite{twofield1}.
In this case the value of the non-Gaussianity parameter is given
by\cite{delta-N, komatsu-s, modulated-inf1}
\begin{equation}
\label{result2}
-\frac{3}{5}f_{nl}=\frac{1}{2}\frac{N_{\cal M M}}{N_{\cal M}^2}
\simeq \frac{\omega''\omega-\omega'^2}{2(\omega')^2 N}.
\end{equation}
Using these results, we will examine the curvature perturbation 
in specific models of inflation.

\subsection{Modulated inflation with a scalar field coupled to gravity 1}
In Ref.\cite{modulated-inflation}, we mentioned that the fluctuation of
the Planck mass may lead to the generation of curvature
perturbation.
In fact, the moduli-dependent Planck mass $M_p({\cal M})$ can be seen as
generalized scalar-tensor theory
(Einstein gravity with a non-minimally coupled massless
scalar field ${\cal M}$),
which leads to the ${\cal M}$-dependence of the kinetic term after
conformal transformations.
Using the above analyses, it is possible to show that 
the isocurvature perturbation related to the modulated Planck scale can
feed the curvature perturbation after horizon crossing.
After conformal transformation, the potential depends on the
scalar field ${\cal M}$ if there is no artificial cancellation.

Let us first examine the conditions for modulated inflation in the
model in which ${\cal M}$ is non-minimally coupled with the scalar
curvature\cite{Tsujikawa-nonm};
\begin{equation}
{\cal L}_g = 
\sqrt{-\hat{g}}\left[\frac{1}{2\kappa^2}-\frac{1}{2}\beta 
\hat{\cal M}^2\right]R.
\end{equation}
The quantity in brackets is positive as far as $\beta \ll 1$ and
$\hat{\cal M} \le M_p$. 
Note that induced coupling to the Ricci scalar may arise from
one-loop gravity corrections.
After conformal transformation 
\begin{equation}
g_{\mu\nu}= \Omega^2 \hat{g}_{\mu\nu}
\end{equation}
with 
\begin{equation}
\Omega^2 = 1-\beta \kappa^2 {\cal M}^2,
\end{equation}
we find the action for the inflaton kinetic term with
\begin{equation}
\omega = \frac{1}{1-\beta \kappa^2 {\cal M}^2}
\end{equation}
and
\begin{equation}
W=\omega^2 V(\phi),
\end{equation}
where ${\cal M}$ is redefined by
\begin{equation}
{\cal M}\equiv\int 
\Omega^{-1}\sqrt{1-(1-6\beta)\beta\kappa^2\hat{\cal M}^2}d\hat{\cal M}.
\end{equation}
The conformal transformation leads to the effective potential
\begin{equation}
W({\cal M},\phi)\simeq  V(\phi)+6 \beta H_I^2{\cal M}^2,
\end{equation}
where we assumed $\beta\kappa^2{\cal M}^2\ll 1$ in the last approximation.
The model has been studied by Tsujikawa et al. in
Ref.\cite{Tsujikawa-nonm} by numerical computation.
However the parameter space where the modulated perturbation is
significant was ``ruled out'', due to the fact that the modulated
perturbation leads to the distortions in inflaton perturbation. 
In fact, in modulated inflation, we consider the situation where
the modulated perturbation dominates the inflaton perturbation,
which may be seen as a ``distortion''.
However, the important point is that the
modulated 
perturbation can lead to a {\bf successful} generation of the CMB
spectra if appropriate conditions are satisfied.
Modulated inflation gives a good approximation when the field ${\cal M}$
satisfies the slow-roll condition.
In this case, the approximation $D\simeq 0$ in Eq.(\ref{ABD}) is good
and there is no strong amplification or suppression
of $\delta {\cal M}$ during inflation.
Using Eq.(\ref{result1}),
we find the condition for the modulated perturbation to dominate
the curvature perturbation to be
\begin{equation}
\left|2N \beta \epsilon^{1/2} \frac{\cal M}{M_p}\right|>1,
\end{equation}
where $|\beta| \ll 1$ and $\delta {\cal M}\simeq \delta \phi$ are
considered. 
If there is no fine-tuning between the slow-roll parameters, 
the spectral index suggests that $|\epsilon| \le |\beta| \sim
O(10^{-2})$ or 
$|\beta| \le| \epsilon|\sim O(10^{-2})$.
We thus conclude that the ratio is at most $N_{\cal M}/N_{\phi}\sim 0.1$
for ${\cal M}\sim M_p$.
Note that the above condition is valid only when there is no
amplification or suppression of the fluctuations $\delta \phi$ or
$\delta {\cal M}$ during inflation.
We next consider fast-roll inflation\cite{fast-roll}
 with $\eta_{\phi\phi}\ge 1$.
 The fluctuation of $\phi$ does not produce classical perturbations when
 $\eta_{\phi\phi}$ is larger than unity. 
For the fast-roll (hybrid) inflation with quadratic potential $\sim
\pm m^2\phi^2/2$, 
the velocity of the inflaton field is given by\cite{fast-roll}
\begin{equation}
\dot{\phi}=-F(\omega) H_I \phi,
\end{equation}
where the function $F(\omega)$ is defined by
\begin{equation}
F(\omega)\equiv \sqrt{\frac{9}{4}\mp\frac{ m^2}{\omega H_I^2}}-\frac{3}{2}.
\end{equation}
Using this result, we find the fluctuation of the number of 
e-foldings to be
\begin{equation}
\delta N \simeq -\frac{F' N}{F}\delta \omega, 
\end{equation}
where the prime denotes the derivative with respect to $\omega$.
Considering the term proportional to $\dot{\phi}^2$,
the slow-roll condition for ${\cal M}$ is 
\begin{equation}
\frac{\omega'}{\omega} \ll \frac{3\sqrt{2}}{M_p \epsilon}.
\end{equation}
In this case the modulated perturbation is the main source of the
curvature perturbation, and the non-Gaussianity parameter $f_{nl}$ 
can be large.

\subsection{Modulated inflation with a scalar field coupled to gravity 2}
The theory with a light scalar field $\hat{\cal M}$
coupled to gravity can be given by the action
\begin{equation}
S=\int d^4 x \sqrt{-g} \left[
f(\hat{\cal M}) R -g(\hat{\cal M}) \left(\nabla \hat{\cal M}\right)^2
-\frac{1}{2} \left(\nabla \phi\right)^2 -V(\phi)\right].
\end{equation}
Considering Jordan-Brans-Dicke theory\cite{JBD-ref}, $f$ and $g$
are written as
\begin{equation}
f=\frac{\hat{\cal M}}{16\pi}, \,\,\,\,\, 
g=\frac{\omega_{BD}}{16\pi \hat{\cal M}},
\end{equation}
where $\omega_{BD}$ is the so-called Brans-Dicke parameter.\footnote{Note that since ${\cal M}$
couples to matter, it will participate to the gravitational sector.
In this case, $M_p$ would not be the value measured in a Cavendish type
experiment. See Ref.\cite{bdconst} for recent study.} 
After a conformal transformation, the action in the Einstein frame is 
given by the action with the inflaton kinetic term with
\begin{equation}
\omega({\cal M})
=\exp\left(-\beta\kappa {\cal M}\right)
\end{equation}
and the potential
\begin{equation}
W=\omega({\cal M})^2 V(\phi).
\end{equation}
Here the dimensionless constant $\beta$ is given by
\begin{equation}
\beta\equiv\sqrt{\frac{2}{2\omega_{BD}+3}} \ll 1,
\end{equation}
and ${\cal M}$ is the field after conformal transformation.
Using Eq.(\ref{result1}), and introducing a new dimensionless parameter
\begin{equation}
\alpha_{\delta} \equiv \frac{\delta \sigma}{\delta {\cal M}},
\end{equation}
we find the condition for modulated inflation:
\begin{equation}
\beta > \frac{\alpha_\delta}{N\sqrt{\epsilon}}.
\end{equation}
These conditions suggest that the modulated perturbation cannot
dominate the 
curvature perturbation for conventional slow-roll inflation.
On the other hand, modulated perturbation can be significant if the
inflaton is fast-rollong and has the parameter $\eta_{\phi\phi}\ge 1$,
as we discussed in the previous subsection.

\section{Conclusions and discussions}
\hspace*{\parindent}
In this paper, we have studied modulated inflation from 
kinetic term. 
We showed explicitly that the analytic result obtained from the evolution
 of the Mukhanov-Sasaki variable is consistent with the 
$\delta N$-formula.  
Using the simple formula obtained in this paper,
we found an analytic condition for modulated inflation.
We also found analytic formula for the non-Gaussianity
 parameter in modulated inflation.

In modulated inflation, the integral of the source term related
to the perturbation of the inflaton velocity leads to the curvature
perturbation. 
We consider modulated inflation to be an 
alternative to conventional inflation, in the sense that it saves
inflation when the conventional inflaton perturbation fail to generate
the CMB spectra.
For example, observation of a large non-Gaussianity parameter (if
confirmed) can be a problem for single-field inflation.
According to Yadav et al\cite{non-gauss}, the WMAP 3-year data may
not favor single field slow-roll inflation.\footnote{Non-gaussianity
may be generated at preheating\cite{fnl_from_PR_other}.} 
Note also that the string $\eta$-problem may prevent a successful
standard single-field inflation scenario.
There are many attempts in this direction.
A curvaton\cite{alt-curvaton} generates the curvature
perturbation long after inflation and saves inflation when the inflaton
perturbation does not lead to the successful generation of the
cosmological perturbation. 
The observation of a low-energy gravitational effect in the Large Hadron
Collider(LHC) may put a strong upper bound for the inflation scale, but
the bound can be evaded in many ways\cite{alt-curvaton-low, alt-hyb-curv}.
Modulated perturbation may lead to inhomogeneous
preheating\cite{alternative-PR} or inhomogeneous
reheating after inflation\cite{alt-inhore}.
Note that inhomogeneous preheating can work with a low inflation scale.
The string $\eta$-problem may be solved by one of these
alternatives\cite{alt-string}.
Note that modulated inflation is consistent with fast-roll inflation.
Moreover, these models are consistent with a large non-Gaussianity.

\section{Acknowledgment}
We wish to thank K.Shima for encouragement, and our colleagues at
Tokyo University for their kind hospitality.


\begin{thebibliography}{1}
\bibitem{modulated-inf1}
 L.~Kofman,
  [arXiv:astro-ph/0303614];
  F.~Bernardeau, L.~Kofman and J.~P.~Uzan,
  ``Modulated fluctuations from hybrid inflation,''
  Phys.\ Rev.\  D {\bf 70}, 083004 (2004)
  [arXiv:astro-ph/0403315];
D.~H.~Lyth,
  ``Generating the curvature perturbation at the end of inflation,''
  JCAP {\bf 0511}, 006 (2005)
  [arXiv:astro-ph/0510443].
\bibitem{alternative-PR}
 T.~Matsuda,
 ``Generating the curvature perturbation with instant preheating,''
  JCAP {\bf 0703}, 003 (2007)
  [arXiv:hep-th/0610232];
T.~Matsuda,
``Cosmological perturbations from inhomogeneous preheating and
	multi-field trapping,''
  JHEP {\bf 0707}, 035 (2007)
  [arXiv:0707.0543 [hep-th]].
\bibitem{modulated-inflation}
 T.~Matsuda,
  ``Modulated Inflation,''
  [arXiv:0801.2648];
T.~Matsuda,
  ``Running spectral index from shooting-star moduli,''
  arXiv:0802.3573 [hep-th].
\bibitem{two-field}
  C. Gordon, D. Wands, B. A. Bassett and R. Maartens,
{\it Adiabatic and entropy perturbations from inflation,
Phys.Rev.D63(2001)023506} [astro-ph/0009131].
\bibitem{alt-string}
T.~Matsuda,
   ``Elliptic inflation: Generating the curvature perturbation without
  slow-roll,''
  JCAP {\bf 0609}, 003 (2006)
  [arXiv:hep-ph/0606137];
D.~H.~Lyth and A.~Riotto,
   ``Generating the curvature perturbation at the end of inflation in
	string theory,''
  Phys.\ Rev.\ Lett.\  {\bf 97}, 121301 (2006)
  [arXiv:astro-ph/0607326];
  T.~Matsuda,
  ``Non-tachyonic brane inflation,''
  Phys.\ Rev.\  D {\bf 67}, 083519 (2003)
  [arXiv:hep-ph/0302035].
\bibitem{reviewer1}
  C.~Ringeval, P.~Brax, C.~van de Bruck and A.~C.~Davis,
  ``Boundary inflation and the WMAP data,''
  Phys.\ Rev.\  D {\bf 73}, 064035 (2006)
  [arXiv:astro-ph/0509727].
\bibitem{Tsujikawa-nonm}
  S.~Tsujikawa and H.~Yajima,
  ``New constraints on multi-field inflation with nonminimal coupling,''
  Phys.\ Rev.\  D {\bf 62}, 123512 (2000)
  [arXiv:hep-ph/0007351];
 A.~A.~Starobinsky, S.~Tsujikawa and J.~Yokoyama,
  ``Cosmological perturbations from multi-field inflation in generalized
  Einstein theories,''
  Nucl.\ Phys.\  B {\bf 610}, 383 (2001)
  [arXiv:astro-ph/0107555].
\bibitem{Lalak-kinmix}
 Z.~Lalak, D.~Langlois, S.~Pokorski and K.~Turzynski,
  ``Curvature and isocurvature perturbations in two-field inflation,''
  JCAP {\bf 0707}, 014 (2007)
  [arXiv:0704.0212].
\bibitem{Bernardeau2002jy}
  F.~Bernardeau and J.~P.~Uzan,
  ``Non-Gaussianity in multi-field inflation,''
  Phys.\ Rev.\  D {\bf 66}, 103506 (2002)
  [arXiv:hep-ph/0207295].
\bibitem{Mukhanov:1997fw}
  V.~F.~Mukhanov and P.~J.~Steinhardt,
  ``Density perturbations in multifield inflationary models,''
  Phys.\ Lett.\  B {\bf 422}, 52 (1998)
  [arXiv:astro-ph/9710038].
\bibitem{Bernardeau:2002jf}
  F.~Bernardeau and J.~P.~Uzan,
  ``Inflationary models inducing non-gaussian metric fluctuations,''
  Phys.\ Rev.\  D {\bf 67}, 121301 (2003)
  [arXiv:astro-ph/0209330].
\bibitem{komatsu-s}
 E.~Komatsu and D.~N.~Spergel,
  ``Acoustic signatures in the primary microwave background bispectrum,''
  Phys.\ Rev.\  D {\bf 63}, 063002 (2001)
  [arXiv:astro-ph/0005036].
\bibitem{delta-N}
  M.~Sasaki and E.~D.~Stewart,
  ``A General Analytic Formula For The Spectral Index Of The Density
  Perturbations Produced During Inflation,''
  Prog.\ Theor.\ Phys.\  {\bf 95}, 71 (1996)
  [arXiv:astro-ph/9507001].
\bibitem{twofield1}
  J.~Garcia-Bellido and D.~Wands,
  ``Metric perturbations in two-field inflation,''
  Phys.\ Rev.\  D {\bf 53}, 5437 (1996)
  [arXiv:astro-ph/9511029];
  J.~Garcia-Bellido and D.~Wands,
  ``Constraints from inflation on scalar - tensor gravity theories,''
  Phys.\ Rev.\  D {\bf 52}, 6739 (1995)
  [arXiv:gr-qc/9506050].
\bibitem{fast-roll}
  A.~Linde,
  ``Fast-roll inflation,''
  JHEP {\bf 0111}, 052 (2001)
  [arXiv:hep-th/0110195];
  K.~Dimopoulos and M.~Axenides,
  JCAP {\bf 0506}, 008 (2005)
  [arXiv:hep-ph/0310194];
T.~Matsuda,
  ``Brane inflation without slow-roll,''
  JHEP {\bf 0703}, 096 (2007)
  [arXiv:astro-ph/0610402].
\bibitem{JBD-ref}
  C.~Brans and R.~H.~Dicke,
  ``Mach's principle and a relativistic theory of gravitation,''
  Phys.\ Rev.\  {\bf 124}, 925 (1961).
\bibitem{alt-curvaton}
 D.~H.~Lyth and D.~Wands,
 ``Generating the curvature perturbation without an inflaton,''
  Phys.\ Lett.\  B {\bf 524}, 5 (2002)
  [arXiv:hep-ph/0110002];
 T.~Moroi and T.~Takahashi,
 ``Effects of cosmological moduli fields on cosmic microwave background,''
  Phys.\ Lett.\  B {\bf 522}, 215 (2001)
  [Erratum-ibid.\  B {\bf 539}, 303 (2002)]
  [arXiv:hep-ph/0110096];
\bibitem{alt-curvaton-low}
 T.~Matsuda,
  ``Curvaton paradigm can accommodate multiple low inflation scales,''
  Class.\ Quant.\ Grav.\  {\bf 21}, L11 (2004)
  [arXiv:hep-ph/0312058];
 K.~Dimopoulos, D.~H.~Lyth and Y.~Rodriguez,
  ``Low scale inflation and the curvaton mechanism,''
  JHEP {\bf 0502}, 055 (2005)
  [arXiv:hep-ph/0411119];
Y.~Rodriguez,
  ``Low scale inflation and the immediate heavy curvaton decay,''
  Mod.\ Phys.\ Lett.\  A {\bf 20}, 2057 (2005)
  [arXiv:hep-ph/0411120];
  T.~Matsuda,
  ``Hilltop Curvatons,''
  Phys.\ Lett.\  B {\bf 659}, 783 (2008)
  [arXiv:0712.2103 [hep-ph]].
\bibitem{alt-hyb-curv}
  T.~Matsuda,
  ``Hybrid Curvatons from Broken Symmetry,''
  JHEP {\bf 0709}, 027 (2007)
  [arXiv:0708.4098 [hep-ph]];
T.~Matsuda,
  ``Topological curvatons,''
  Phys.\ Rev.\  D {\bf 72}, 123508 (2005)
  [arXiv:hep-ph/0509063].

\bibitem{alt-inhore}
 G.~Dvali, A.~Gruzinov and M.~Zaldarriaga,
``A new mechanism for generating density perturbations from inflation,''
  Phys.\ Rev.\  D {\bf 69}, 023505 (2004)
  [arXiv:astro-ph/0303591];
 K.~Enqvist, A.~Mazumdar and M.~Postma,
   ``Challenges in generating density perturbations from a fluctuating
	inflaton coupling,''
  Phys.\ Rev.\  D {\bf 67}, 121303 (2003)
  [arXiv:astro-ph/0304187];
F.~Vernizzi,
  ``Cosmological perturbations from varying masses and couplings,''
  Phys.\ Rev.\  D {\bf 69}, 083526 (2004)
  [arXiv:astro-ph/0311167].
\bibitem{non-gauss}
 A.~P.~S.~Yadav and B.~D.~Wandelt,
``Detection of primordial non-Gaussianity (fNL) in the WMAP 3-year data at
  above 99.5
  arXiv:0712.1148 [astro-ph].
\bibitem{bdconst}
  V.~Acquaviva and L.~Verde,
  ``Observational signatures of Jordan-Brans-Dicke theories of gravity,''
  JCAP {\bf 0712}, 001 (2007)
  [arXiv:0709.0082 [astro-ph]].
\bibitem{fnl_from_PR_other}
K. Enqvist, A. Jokinen, A. Mazumdar, T. Multamaki and A. Vaihkonen,
{\it Non-Gaussianity from Preheating, Phys.Rev.Lett.94(2005)161301}
[astro-ph/0411394];
K. Enqvist, A. Jokinen, A. Mazumdar, T. Multamaki and A. Vaihkonen,
{\it Non-gaussianity from instant and tachyonic preheating,
JCAP {0503}, 010 (2005)} [hep-ph/0501076];
A. Jokinen and A. Mazumdar,
{\it Very Large Primordial Non-Gaussianity from multi-field: Application  
to Massless Preheating, JCAP {0604}, 003 (2006)} [astro-ph/0512368].
\end{thebibliography}
\end{document}